\documentclass[preprin,showpacs,preprintnumbers,eqsecnum,amsmath,amssymb]{revtex4}
\usepackage{graphics,epsfig}
\usepackage{graphicx}
\usepackage{dcolumn}
\usepackage{bm}

\begin{document}
\baselineskip=0.8 cm
\title{Dispersion relation and surface gravity of universal horizons}

\author{Chikun Ding${}^{a,b}$ }
\thanks{ Email: Chikun\_Ding@huhst.edu.cn; dingchikun@163.com}
\author{Changqing Liu${^a}$}
  \affiliation{$^{a}$ Department of Physics, Hunan University of Humanities, Science and Technology, Loudi, Hunan
417000, P. R. China\\
$^{b}$Key Laboratory of Low Dimensional
Quantum Structures and Quantum Control of Ministry of Education,
and Synergetic Innovation Center for Quantum Effects and Applications,
Hunan Normal University, Changsha, Hunan 410081, P. R. China}

\vspace*{0.2cm}
\begin{abstract}
\baselineskip=0.6 cm
\begin{center}
{\bf Abstract}
\end{center}

In Einstein-aether theory, violating Lorentz invariance permits some super-luminal communications, and the universal horizon can trap excitations traveling at arbitrarily high velocities. To better understand the nature of these universal horizons, we first modify the ray tracing method, and then use it to study their surface gravity in charged Einstein-aether black hole spacetime. Instead of the previous result in Ref. [Phys. Rev. D 89, 064061], our results show that the surface gravity of the universal horizon is dependent on the specific dispersion relation, $\kappa_{UH}=2(z-1)\kappa_{uh}/z$, where $z$ denotes the power of the leading term in the superluminal dispersion relation, characterizing different species of particles. And the associated Hawking temperatures also are different with $z$. These findings, which coincide with those in Ref. [Nucl. Phys. B 913, 694] derived by the tunneling method, provide some full understanding of black hole thermodynamics in Lorentz-violating theories.

\end{abstract}


 \maketitle

\vspace*{0.2cm}
\section{Introduction}
Invariance under Lorentz transformation is one of the pillars of both Einstein's General Relativity (GR) and the Standard Model of particle physics.  However, Lorentz invariance may not be an exact symmetry at all energies  \cite{mattingly}. Condensed matter physics, which has an analog of Lorentz invariance (LI), suggests some scenarios of Lorentz violation (LV): a) LI is an approximate symmetry emerging at low energies and violated at ultrahigh energies \cite{chadha}; b) LI is fundamental but broken spontaneously \cite{klinkhamer}.
Any effective description must break down at a certain cutoff scale, which signs the emergence of new physical degrees of freedom beyond that scale. Examples of this include the hydrodynamics, Fermi's theory of beta decay \cite{bhattacharyya2} and quantization of GR \cite{shomer} at energies beyond the Planck energy. Lorentz invariance also leads to divergences in quantum field theory which can be cured with a short distance of cutoff that breaks it \cite{jacobson}. This conjecture seems to have been proven by astrophysical observations on high-energy cosmic rays \cite{abdo}.
Additionally, LV may be relatively large but hidden from the many experiments to date \cite{kostelecky2009}.

 Thus, the study of LV is a valuable tool to probe the foundations of GR without preconceived notions of the numerical sensitivity \cite{shao}. These studies include LV in the neutrino sector \cite{dai}, the Standard-Model Extension \cite{colladay}, LV in the non-gravity sector \cite{coleman}, and LV effect on the formation of atmospheric showers \cite{rubtsov}.
A more recent area for searching for LV is in the pure gravity sector, such as gravitational Cerenkov radiation \cite{kostelecky2015} and gravitational wave dispersion \cite{kostelecky2016}.
Einstein-aether theory can be considered as an effective description of Lorentz symmetry breaking in the gravity sector and has been extensively used in order to obtain quantitative constraints on Lorentz-violating gravity\cite{jacobson2}. On another side, violations of Lorentz symmetry have
been used to construct modified-gravity theories that account
for dark-matter phenomenology without any actual dark mater \cite{bekenstein2004}.

There are several gravitational theories that violate the Lorentz invariance  \cite{mattingly}, e.g., Ho\v{r}ava-Lifshitz theory \cite{Horava}, ghost condensations \cite{Shinji}, warped brane worlds, and
Einstein-aether theory \cite{jacobson2}  which originated from the scalar-tensor theory \cite{gong}. In Einstein-aether theory, the background tensor fields break the Lorentz symmetry only down to a rotation subgroup by the existence of a preferred time direction at every point of spacetime, i.e., existing a preferred frame of reference established by aether vector $u^a$. The introduction of the aether vector allows for some novel effects, e.g., matter fields can travel faster than the speed of light \cite{jacobson3}, new gravitational wave polarizations can spread at different speeds \cite{jacobson4}. It is the
 universal horizons that can trap excitations traveling at arbitrarily high velocities. The superluminal interactions might allow us to probe the ``behind" of black hole and cosmological horizons, normally off limits by relativistic causality \cite{sundrum}. This aether theory can
also be considered as  a realization of dynamic self-interaction of complex systems moving with a spacetime dependant macroscopic velocity. As to an accelerated expansion of the universe, this dynamic self-interaction can produce the same cosmological effects as the dark energy \cite{balakin2}.

There are many studies on the universal horizon up to date \cite{tian}.  Black hole thermodynamics provides an alternative way to probe the ultraviolet (UV) physics of LV models. Some authors claim that the universal horizon may radiate thermally at a fixed temperature and strengthen a possible thermodynamic interpretation though there is no universal light cone \cite{berglund2} (See also \cite{michel} for a different suggestion).
Berglund {\it et al} \cite{berglund} used tunneling method to study black radiation at the universal horizon in the special choice of $z=2$, where $z$ denotes the power of the leading term in the nonlinear dispersion relation, characterizing different species of particles. They found that the universal horizon radiates as a blackbody at a fixed temperature though the scalar fields violate local Lorentz invariance. Another different viewpoint is in \cite{michel}, in which  the late time radiation was computed. Cropp {\it et al} \cite{cropp} studied ray tracing, found the evidence of Hawking radiation at universal horizon and ray lingering near Killing horizon and, gave the covariant definition of surface gravity of universal horizon. However, it is incorrect that the surface gravity they obtained is independent of the specific dispersion relations, see Appendix for more details.

In Ref. \cite{ding2}, we have used the tunneling method to study Hawking radiation of the charged Einstein-aether black hole spacetime.  Our
results at the Killing horizons confirm the previous ones, i.e., at high frequencies the corresponding radiation remains thermal and the nonlinearity
of the dispersion does not alter the  Hawking radiation significantly.  On the contrary, superluminal particles with $z>1$ are only   created at universal horizons and are
radiated out to infinity. Although the radiation is also thermal spectrum, different species of particles in general experience different temperatures that loose their universal meaning. It is shown that this LV model is not capable of reproducing the success of black hole thermodynamics in GR, and may suggest some hints on necessary properties of would-be UV completion.

In this paper, we reconsider ray tracing \cite{cropp} in Einstein-aether black hole spacetime, and find that, instead of the results of Cropp {\it et al}, the surface gravity of the universal horizon is indeed dependent on the power $z$, hence for the associated Hawking temperature. These findings may be important for the full understanding Lorentz-violating theories.  The rest of the paper is organized as follows. In Sec. II we review the peeling-off surface gravity derived in an usual static black hole spacetime. In Sec. III we modify this peeling-off method and extend it to the universal horizon. In Sec. IV, we consider a general superluminal dispersion relation with power $z$ and derive its group velocity. In Sec. V, we give the peeling-off surface gravity of a universal horizon and the relation between it and the covariant surface gravity. In Sec. VI, we present our main conclusions.

\section{Rays peeling and surface gravity of an usual static black hole}

In this section, we first review the standard description of particle trajectories propagating in an usual static spacetime, with particular attention on their behavior close to the Killing horizon.
The most usual textbook notion of surface gravity $\kappa_i$ is defined by the ``inaffinity" of the naturally normalized null geodesics on the horizon of a static spacetime \cite{wald},
\begin{eqnarray}
 \chi^a\nabla_a\chi^b=\kappa_i\chi^b,
\end{eqnarray}
where $\chi^a$ is the Killing vector. The other conception of surface gravity relates to the peeling off properties of null geodesics near the horizon. Since for stationary Killing horizon these two notions coincide, then in this paper, we pay our attention to this ``peeling" surface gravity.

The static black hole metric reads
\begin{eqnarray}
 ds^2=-e(r)dt^2+\frac{1}{e(r)}dr^2+r^2(d\theta^2+\sin^2\theta d\varphi^2).
\end{eqnarray}
While the spacetime is perfectly well behaved there, the coordinates $(t,r)$ become singular at the horizon $r=r_H$, and no longer in a one-to-one correspondence with spacetime events. For the sake of constructing spacetime diagram, one usually changes it into Eddington-Finkelstein coordinates
\begin{eqnarray}\label{EF}
 ds^2=-e(r)dv^2+2dvdr+r^2(d\theta^2+\sin^2\theta d\varphi^2),
\end{eqnarray}
with the transformation ($v$ is the advanced time)
\begin{eqnarray}
 dv=dt+\frac{dr}{e(r)}.
\end{eqnarray}
It can be verified that the $(v,r)$ coordinates cover regions {\bf I} and {\bf II} of the Kruskal diagram.

The ingoing light rays move with $dv=0$, that is, the coordinate lines can be oriented at 45 degrees. The outgoing rays will be given by
\begin{eqnarray}\label{er}
\frac{dt}{dr}=\frac{1}{2}\frac{dv}{dr}=\frac{1}{e(r)}.
\end{eqnarray}
If the geodesic is near the metric horizon $e(r_H)=0$, then we can Taylor expand it as
\begin{eqnarray}
\frac{dr}{dt}=\frac{de(r)}{dr}\Big|_{r_H}(r-r_H)+\mathcal{O}(r-r_H)^2.
\end{eqnarray}
Defining a peeling surface gravity
\begin{eqnarray}\label{kh}
\kappa_{p}=\frac{1}{2}\frac{de(r)}{dr}\Big|_{r_H}=\kappa_{KH},
\end{eqnarray}
where $\kappa_{KH}$ is the the standard surface gravity  at the the Killing horizon in the case of static black hole, then we have
\begin{eqnarray}
\frac{dr}{dt}=2\kappa_p(r-r_H)+\mathcal{O}(r-r_H)^2.
\end{eqnarray}
For two null geodesics $r_1(t)$ and $r_2(t)$ on the same side of the evolving horizon \cite{cropp2}
\begin{eqnarray}
\frac{d|r_1-r_2|}{dt}\approx2\kappa_p|r_1(t)-r_2(t)|.
\end{eqnarray}
This makes manifest the fact that $\kappa_p$ is related to the exponential peeling off properties of null geodesics near the horizon. It is this
quantity $\kappa_p$ that we have seen is ultimately connected to the temperature of the Hawking flux \cite{barcelo}.

\section{Extension to Einstein-aether spacetime}
In this section, we analyze the motion of particles endowed with Lorentz-violating dispersion relations and  then we construct the ray trajectories in an aether black hole spacetime. This modified dispersion relations arise due to the interaction of particles with the aether field. So the trajectories are not simply the geodesics determined by the given metric.

The static spherically symmetric spacetime in Einstei-aether theory takes the Eddington-Finklestein coordinates form (\ref{EF})
and
\begin{equation}\label{e(r)}e(r)=(u\cdot\chi)^2-(s\cdot\chi)^2.\end{equation}
The corresponding timelike Killing vector, aether vector and its orthogonal one are, respectively,
\begin{equation}\label{us}
  \chi^a=(1,0,0,0),\quad u^a=\big(\alpha,~\beta,~0,~0\big),\quad s^a=\big(\alpha,~\beta+\frac{1}{\alpha},~0,~0\big),
\end{equation}
where $\alpha(r)$ and $\beta(r)$ are  functions of $r$ only
\begin{eqnarray}\label{abe}
 \alpha(r)=\frac{1}{(s\cdot\chi)-(u\cdot\chi)},\quad \beta(r)=-(s\cdot\chi).
\end{eqnarray}
 Then,  the metric can be written as  $g_{ab}=-u_au_b+s_as_b+\hat{g}_{ab}$, where $\hat{g}_{ab}$ is projection tensor onto the spatial two-sphere surface and we have the
 constraints $u^2=-1, ~s^2=1,~ u\cdot s=0$.

In this black hole spacetime, the aether time gives the aether one-form
\begin{eqnarray}
ds_u=u_adx^a=u_vdv+u_rdr=u_vd\tau,\end{eqnarray}with\begin{eqnarray} \;\tau=v+\int\frac{u_r}{u_v}dr,
\end{eqnarray}
which splits spacetime into a great number of spacelike slices.
There are also  some peeling-like behaviors of constant $\tau$ slices near the universal horizon.
So that a notion of $\kappa_p$ can be associated to these slices.
One can calculate this surface gravity via the theory of Hamiltonian as a dispersion relation in geometrodynamics \cite{cropp,misner}.
The variational principle gives Hamilton's equations for the rates of change
\begin{eqnarray}\label{hamilton}
\frac{dx^\alpha}{d\lambda}=\frac{\partial \mathcal{H}}{\partial k_\alpha},\;\frac{dk_\beta}{d\lambda}=-\frac{\partial \mathcal{H}}{\partial x^\beta},
\end{eqnarray}
which show that $\mathcal{H}$ must be a constant, independent of the time-like $\lambda$ along the trajectory of the particle.  Its value has to be imposed as an initial condition, $\mathcal{H}=0$. The four momentum vector must be thrown in the aether frame
\begin{eqnarray}
k_a=-k_uu_a+k_ss_a,
\end{eqnarray}
where $k_u=(u\cdot k)=u^ak_a,~k_s=(s\cdot k)$ are the aether frame energy and momentum.

Suppose that the dispersion relation is $\omega=f(k_s)$, where $\omega=-k_u$,  then the Hamiltonian can be chosen as $\mathcal{H}=[\omega^2-f^2(k_s)]/2$. Substituting it into the above equation (\ref{hamilton}), one can obtain
\begin{eqnarray}
\frac{dt}{dr}=\frac{1}{2}\frac{dv}{dr}=\frac{1}{2}
\frac{u^v+v_gs^v}{u^r+v_gs^r}
\end{eqnarray}
for outgoing particles, where $v_g=\partial \omega/\partial k_s$ is the group velocity of the propagating particles.
It is just the trajectory for an instantaneously propagating ray.
Taylor expanding the trajectory near the universal horizon $(u\cdot\chi)=0$, we obtain
\begin{eqnarray}
\frac{dr}{dt}=2\kappa_{{UH}}(r-r_{UH})
+\mathcal{O}(r-r_{UH})^2.
\end{eqnarray}
Then, we can define the surface gravity at the universal horizon via this peeling one,
\begin{eqnarray}
\kappa_{UH}\equiv\frac{1}{2}\frac{d}{dr}\frac{dr}{dt}\Big|_{{UH}}
=\frac{d}{dr}\frac{u^r+v_gs^r}{u^v+v_gs^v}\Big|_{{UH}}.
\end{eqnarray}
As soon as the modified dispersion relation is given, one can find the group velocity and, lastly obtain the surface gravity of the universal horizon of a given Einstein-aether black hole.

It is interesting that this result can reduce to the usual one (\ref{kh}). If $v_g=1$, the group velocity for the luminal particles, by using Eqs. (\ref{e(r)}-\ref{abe}), the above equation can be reduced to
\begin{eqnarray}
\kappa_{KH}=\frac{1}{2}\frac{de(r)}{dr}\Big|_{{r_{KH}}},
\end{eqnarray}
which is the same as (\ref{kh}) at the Killing horizon $e(r_{KH})=0$.

\section{Dispersion relation and group velocity}

In this paper, we consider the superluminal dispersion relation, which is given by  \cite{unruh,WWM},
 \begin{eqnarray}\label{ksquare}
(-k_u)^2=k_0^2\sum_{n=1}^{z}
a_n\Big(\frac{k_s}{k_0}\Big)^{2n},
\end{eqnarray}
where  $a_n$'s are greater than zero and dimensionless constants, which will be considered as the order of unit in the following discussions \cite{WWM}, and $z$ is an integer. Lorentz symmetry requires that
$(a_1, z)= (1, 1)$. Therefore, in this paper we shall set $a_1 = 1$.
The $k_0$ is the UV Lorentz-violating scale for the matter \cite{cropp} or the suppression mass scale \cite{yagi}. The experimental viable range for the $k_0$ is rather broad and its value shows the size of Lorentz violation of the given field.

The Killing energy is
\begin{eqnarray}
\Omega=-(k\cdot\chi)=-k_u(-u\cdot\chi)-k_s(s\cdot\chi).
\end{eqnarray}
Near the universal horizon, $k_s$ diverges there and can be parameterized by
\begin{eqnarray}\label{ks}
k_s=\frac{k_0b(r)}{(-u\cdot\chi)^m},
\end{eqnarray}
where $b(r)$ is regular and nonzero at the horizon and, $m$ is a constant to be determined. We have
\begin{eqnarray}
(-u\cdot\chi)^{mz-1}\Omega=k_0\sqrt{a_z}b^z-k_0b(-u\cdot\chi)^{m(z-1)-1}
(s\cdot\chi).
\end{eqnarray}
Hence, $m$ and $b$ can be read off
\begin{eqnarray}
m=\frac{1}{z-1},\;\sqrt{a_z}b^{z-1}(r_{UH})=
(s\cdot\chi)|_{UH}.
\end{eqnarray}
From (\ref{ksquare}) and (\ref{ks}), the group velocity is
\begin{eqnarray}
v_g=-\frac{\partial k_u}{\partial k_s}\approx z\sqrt{a_z}\Big(\frac{k_s}{k_0}\Big)^{z-1}_{UH}=\frac{z(s\cdot\chi)|_{UH}}{(-u\cdot\chi)}.
\end{eqnarray}
One can see that the group velocity $v_g\rightarrow\infty$ at the universal horizon $(u\cdot\chi)(r_{UH})=0$, and is dependent on the specific form of dispersion relation, i.e., the parameter $z$.

The trajectory
\begin{eqnarray}
\frac{dv}{dr}&=&
\frac{u^v+v_gs^v}{u^r+v_gs^r}=
\frac{1}{(s\cdot\chi)-(u\cdot\chi)}\cdot
\frac{1+v_g}{-(s\cdot\chi)-(u\cdot\chi)v_g}\nonumber\\
&\approx&\frac{z}{z-1}\frac{1}{(s\cdot\chi)|_{UH}(-u\cdot\chi)},
\end{eqnarray}
near the universal horizon $(u\cdot\chi)=0$.
The surface gravity at the universal horizon is \footnotemark\footnotetext{This result is similar to the formula (3.56) with inverse sign given in \cite{cropp2016}. However, its formula (3.58) isn't the correct surface gravity, but the normal one which is given in our Eq. (\ref{imuh}).}
\begin{eqnarray}\label{kuh}
\kappa_{UH}=\frac{1}{2}\frac{d}{dr}\frac{dr}{dt}\Big|_{{UH}}
=\frac{z-1}{z}(s\cdot\chi)\frac{d(-u\cdot\chi)}{dr}\Big|_{UH},
\end{eqnarray}
which depends on the specific dispersion relation. So the associated black hole temperature looses its universal meaning for all species of particles, then such black hole would therefore not be in equilibrium and hence, conflicts with the ``zeroth law" and the generalized second law \cite{eling}.

\section{Surface gravity of a Charged Einstein-aether black hole}

In this section we can calculate the surface gravity at the universal horizon of a charged Einstein-aether black hole \cite{ding}.
The static spherically symmetric spacetime in Einstein-Maxwell-aether theory takes the Eddington-Finklestein coordinates form (\ref{EF}).
When coupling constants $c_{14}=0$ ($c_{14}\equiv c_1+c_4$, respectively), \begin{eqnarray}
 &&(s\cdot\chi)=\frac{1}{\sqrt{\mathrm{g}}}\frac{r_{UH}^2}{r^2}
 \sqrt{\frac{1}{3}\Big(1-\frac{Q^2}{r_{UH}^2}\Big)}
 ,\quad r_0=\frac{2}{3}r_{UH}\Big(2+\frac{Q^2}{r_{UH}^2}\Big),\nonumber\\
 &&(u\cdot\chi)=-(1-\frac{r_{UH}}{r})\sqrt{1+\frac{r_{UH}}{3r}
 \left(1-\frac{Q^2}{r^2_{UH}}\right)\left(2+\frac{r_{UH}}{r}\right)}~,
\end{eqnarray}
where $\mathrm{g}\equiv1-c_{13}$, $r_0$ is a parameter related to the total mass of the aether black holes, and $r_{UH}$ is the location of the universal horizons.
When coupling constants $c_{123}=0$, \begin{eqnarray}
 &&(u\cdot\chi)=-1+\frac{r_{UH}}{r}
 ,\quad r_0=2r_{UH},\nonumber\\
 &&(s\cdot\chi)=\frac{r_{UH}}{r}\sqrt{\frac{\mathrm{p}}{\mathrm{g}}-
 \frac{Q^2}{\mathrm{g}r^2_{UH}}}~,
\end{eqnarray}
where $\mathrm{p}\equiv1-c_{14}/2$.
We apply the aether coupling constants condition $0<c_{14}<2,~ 2+c_{13}+3c_2>0,~ 0\leq c_{13}<1$.

Putting everything together in (\ref{kuh}), we finally obtain the surface gravity
for the case of $c_{14}=0$,
\begin{eqnarray}
 &&\kappa_{UH}=\frac{z-1}{z \sqrt{3\mathrm{g}}r_{UH}}\sqrt{(1-\frac{Q^2}{r_{UH}^2})
 (2-\frac{Q^2}{r_{UH}^2})}\;.
\end{eqnarray}
If $z=2$ and $Q=0$, it becomes the result in Ref. \cite{berglund2}, $T_{UH}=\kappa_{UH}/2\pi=\sqrt{2}/\big[4\pi r_{UH}\sqrt{3\mathrm{g}}~\big]$, which was obtained in Painlev\'{e}-Gullstrand (PG) coordinates.
In the case of $c_{123}=0$, the surface gravity is
\begin{eqnarray}
 &&\kappa_{UH}=\frac{z-1}{z r_{UH}}\sqrt{\frac{\mathrm{p}}{\mathrm{g}}-
 \frac{Q^2}{\mathrm{g}r^2_{UH}}}~.
\end{eqnarray}
If $z=2$ and $Q=0$, it becomes the result in Ref. \cite{berglund2}, $T_{UH}=\kappa_{UH}/2\pi=\sqrt{\mathrm{p}}/\big[4\pi r_{UH}\sqrt{\mathrm{g}}~\big]$.

Now we consider the comparison of this peeling-off surface gravity to a covariant definition one, which is the normal derivative to the horizon of the redshift factor. For the Killing horizon, the redshift factor is $\chi^2$. However for the universal horizon it is $(u\cdot\chi)$ which is constant on the universal horizon and captures the role of the aether in the propagation of the physical rays \cite{cropp}. This definition is
\begin{eqnarray}\label{imuh} \kappa_{uh}=\frac{1}{2}u^a\nabla_a(u\cdot\chi)\Big|_{UH}
=\frac{1}{2}(s\cdot\chi)\frac{d(-u\cdot\chi)}{dr}\Big|_{UH}.
\end{eqnarray}
The relation between both definitions is
 \begin{eqnarray}\label{uh}
 \kappa_{UH}=\frac{2(z-1)}{z} \kappa_{uh}
\end{eqnarray}
 One can see that the peeling-off surface gravity $\kappa_{UH}$ isn't equal to the normal surface gravity $\kappa_{uh}$ in general. But it is exactly equal to the associated temperature $T_{UH}=\kappa_{UH}/2\pi$ derived in Ref. \cite{ding2} by the tunneling method\footnotemark\footnotetext{In Einstein-aether spacetime, the geodesics of the peeling particles aren't the usual null geodesic \cite{wei2016} any more. The coincidence between the peeling-off surface gravity and the tunneling temperature, shows that the radiating particles also do not move along the usual null geodesics any longer. }. It is shown that there are Hawking radiation associated with the universal horizon. However, this radiation may be sensitive to the details of the UV completion, and it is similar to the Hawking radiation in the ghost condensate \cite{dubovsky}. It maybe show that the standard black hole thermodynamics will break down in the presence of LV, and the UV completed theory should be very unusual.

\section{Summary}

In Einstein-aether theory, violating Lorentz invariance permits some super-luminal communications, and the universal horizon can trap excitations traveling at arbitrarily high velocities. The trajectories of these super-luminal rays are no longer metric geodesics due to the presence of the aether. In this paper we consider rays' peeling behaviors in the charged Einstein aether black hole spacetime. We first review the peeling-off surface gravity derived by Cropp {\it et al} \cite{cropp} and, have found that there is a mathematical error (in Appendix). We then modify this process in Einstein-aether spacetime and find that it can reduce to the usual result (\ref{kh}) for luminal particles, whose group velocity $v_g=1$.

By using the superluminal dispersion relation, we apply this improved peeling-off surface gravity to the charged Einstein-aether black hole. Instead of the previous result in Ref. \cite{cropp}, our results show that the surface gravity of the universal horizon is indeed dependent on the specific dispersion relation, $\kappa_{UH}=2(z-1)\kappa_{uh}/z$, where $\kappa_{uh}$ is the normal surface gravity (\ref{imuh}), $z$ denotes the power of the leading term in the nonlinear dispersion relation, characterizing different species of particles. And the associated Hawking temperatures also are different with $z$. These findings coincide with those in Ref. \cite{ding2} derived by the tunneling method, and are similar to the Hawking radiation in the ghost condensate \cite{dubovsky}.

So the black hole temperature looses its universal meaning for all species of particles. Hence such black hole would therefore not be in equilibrium and, conflicts with the ``zeroth law" and the generalized second law \cite{eling}. It shows the standard black hole thermodynamics will be broken down in the presence of LV, and the UV completed theory should be very unusual. These finds
 provide some full understanding of black hole thermodynamics in Lorentz-violating theories.

{\it Note added.}---Nine days before this article was published in the archive (arXiv), the work \cite{cropp2016} appeared online in the same database, containing a partial overlap with our work.

\appendix\section{Review of the incorrect trajectory}
In Ref. \cite{cropp}, the authors calculate the value of surface gravity of the universal horizon which is the relevant one for the rays propagating with infinite group velocity $v_g$. However, there is a mathematical error in their deriving process which ultimately leads to an incorrect result. We firstly review their original process.

Generically any particles propagating in the Einstein aether spacetime will have a four-velocity that can be given in the orthogonal frame provided by $u^a$ and $s^a$ as
\begin{eqnarray}
 V^a=u^a+v_gs^a.
\end{eqnarray}
The trajectory for an instantaneously propagating ray would then be given by
\begin{eqnarray}\label{traject}
 \frac{dv}{dr}=\frac{V^v}{V^r}=\lim_{v_g\rightarrow\infty}\frac{u^v+v_gs^v}
 {u^r+v_gs^r}=\frac{s^v}
 {s^r}.
\end{eqnarray}
The group velocity $v_g$ is dependent on the specific form of dispersion relation $\omega=f(k_s)$, and $v_g=\partial\omega/\partial k_s=f'(k_s)$. So one can see that above trajectory is indeed independent of the specific superluminal dispersion.
Then the surface gravity of the universal horizon is
\begin{eqnarray}\label{result}
\kappa_{uh}=\frac{1}{2}\frac{d}{dr}\Big(\frac{s^v}{s^r}\Big)\Big|_{uh},
\end{eqnarray}
which is also independent of the specific superluminal dispersion relation.

But, there has a mathematical error in (\ref{traject}). In the numerator, both $u^v$ and $s^v$ are finite and nonzero at the universal horizon, so the term $u^v$ can be ignored in the limit $v_g\rightarrow\infty$. However, in the denominator, $s^r$ is zero at the universal horizon, therefore the product of $v_gs^r$ maybe finite and $u^r$ {\it cannot be ignored} ! In their paper, $u^r,v_g$, and $s^r$ are given by, respectively,
\begin{eqnarray}
u^r=-(s\cdot\chi),\;s^r=(-u\cdot\chi),\;v_g=1+3l^2k_s^2
\sim1+3\frac{(s\cdot\chi)|_{uh}}{(u\cdot\chi)}.
\end{eqnarray}
Hence, the product is
\begin{eqnarray}
v_gs^r=(-u\cdot\chi)-3(s\cdot\chi)|_{uh},
\end{eqnarray}
which is indeed finite at the universal horizon $(u\cdot\chi)=0$ and, the quantity $u^r$ {\it cannot be ignored}. Ultimately the result (\ref{result}) is incorrect.

\begin{acknowledgments}
 We would like to thank Stefano Liberati for supporting this work. C.D. was supported by the National Natural Science Foundation
of China (grant No. 11247013), Hunan Provincial Natural Science Foundation of China (grant No. 2015JJ2085), and the fund under grant No. QSQC1203. C.L. was supported by the special fund of National Natural Science Foundation
of China (grant No. 11447168).

\end{acknowledgments}

\end{document}